\documentclass[prl,amssymb,twocolumn,tightenlines,showpacs]{revtex4-1}

\usepackage{graphicx}
\usepackage{verbatim}
\usepackage{amsmath}

\begin{document}


\title{Realizing a variable isotropic depolarizer}


\author{A. Shaham}
\affiliation{Racah Institute of Physics, Hebrew University of Jerusalem, Jerusalem 91904, Israel}
\author{H.S. Eisenberg}
\affiliation{Racah Institute of Physics, Hebrew University of Jerusalem, Jerusalem 91904, Israel}



\begin{abstract}
We demonstrate an isotropic depolarizing channel with a controllable
degree of depolarization. The depolarizer is composed of four
birefringent crystals and half-wave plates. Quantum process
tomography results of the depolarization effect on single photons
agree well with the theoretical prediction. This depolarizer can be
used to test quantum communication protocols with photons.
\end{abstract}


\maketitle

The study of light depolarization is fundamental to the field of
optics. Optical devices that completely depolarize light are
desirable when polarization dependent effects need to be eliminated
\cite{Bohm81,Wang99}. The Lyot depolarizer, which was invented more
than eighty years ago, is an example for such a device.\cite{Lyot}
In the last 15 years, many experiments have encoded quantum
information in the polarization of single photons
\cite{Mattle96,Gisin02}. The polarization of each photon represents
a two-state quantum system, a qubit. The depolarization process is
interpreted as channel induced noise which results in decoherence of
the photonic qubit. In order to study the effect of noise on quantum
information channels, there is a need for a physical realization of
controlled depolarization processes. Previously, photonic noisy
channels which partially depolarize light were implemented in
several ways
\cite{Kwiat_Dephasing,Puentes_Wedge,Almeida_ESD,Karpinski,Shaham,Chiuri}.
Here, we report on the implementation of a simple controllable
isotropic depolarizer, which depolarizes to the same extent light of
any initial polarization, to any required final degree of
polarization.

The Lyot depolarizer is composed of two birefringent crystals, where
one crystal is twice as long as the other and their principal axes
are oriented at $45^\circ$ with respect to each other. The Lyot
depolarizer completely depolarizes light whose coherence length
$t_c$ is shorter than the polarization temporal walk-off $\tau$ in
the shorter crystal:
\begin{equation}\label{temp_delay}
t_{c}<\tau= L\frac{\Delta n}{c}\,,
\end{equation}
where $L$ is the length of the shorter crystal, $\Delta n$ is its
birefringent refractive index difference and $c$ is the speed of
light. The birefringent crystals couple between the polarization and
the temporal degrees of freedom: every crystal entangles the
polarization degrees of freedom with two different temporal modes.
Depolarization occurs because photon detection can not resolve
between the different temporal modes. The addition of more crystals
will further increase the number of temporal modes. Adding
wave-plates before the crystals and between them (or rotating the
crystals) affects every temporal mode separately, and adds more
possibilities to the depolarization process. The combined
depolarization process of the crystals and the wave-plates can be
explicitly calculated for any configuration \cite{Shaham}.

The polarization state of a single photon, as well as of classical
light, can be described either by its density matrix $\hat{\rho}$,
or by a point in the Poincar\'{e} sphere representation. The
Cartesian coordinates of this point are the Stokes parameters
$\{S_1, S_2, S_3\}$, where $S_0\equiv1$. Here we use the convention
that $S_1$ represents the linear horizontal and vertical
polarizations $|H\rangle$ and $|V\rangle$, $S_2$ the linear diagonal
polarizations $|P\rangle=(|H\rangle+|V\rangle)/\sqrt{2}$,  and
$|M\rangle=(-|H\rangle+|V\rangle)/\sqrt{2}$, and $S_3$ the circular
polarizations $|R\rangle=(|H\rangle+i|V\rangle)/\sqrt{2}$ and
$|L\rangle=(i|H\rangle+|V\rangle)/\sqrt{2}$). The length of the
Stokes vector $D=\sqrt{S^2_1+S^2_2+S^2_3}$ is the state's degree of
polarization. For polarized states $D=1$, whereas for partially
polarized states $0<D<1$. The center of the Poincar\'{e} sphere
$D=0$ represents the completely unpolarized state. Characterization
of the polarization state is performed by several projection
measurements using the Quantum State Tomography (QST) procedure
\cite{Kwiat_Tomo}.

Given a depolarizing configuration, its effect on any input
polarization state $\hat{\rho}$ can be described by the mapping
$\hat{\rho}'=\mathcal{E}({\hat{\rho}})$ . The $\mathcal{E}$ map can
be uniquely described by the elements of the positive Hermitian
process matrix $\chi$:
\begin{equation}\label{process matrix}
\mathcal{E}(\hat{\rho})=\sum_{m,n}\chi_{mn}\hat{E}_{m}\hat{\rho}\hat{E}_{n}^\dag\,,
\end{equation}
where  $\hat{E}_m$ are matrices that span the vector space of
$\hat{\rho}$. Assuming the channel has no dissipation (e.g. light
intensity is preserved), the $\chi$ matrix satisfies
$\textrm{Tr}\{\chi\}=1$. The elements of the $\chi$ matrix can be
experimentally determined by a Quantum Process Tomography (QPT)
procedure which requires several QST measurements \cite{Chuang}.
When no depolarization occurs, $\chi$ has one eigenvalue that equals
1, and the rest are zeros. If two or more eigenvalues of $\chi$
differ from zero, the channel depolarizes light. Isotropic
depolarization occurs when the degree of polarization of the output
state is equal for any initial polarized state. Such a process is
described by a matrix with three nonzero equal eigenvalues. A
complete depolarization occurs when all four eigenvalues of $\chi$
are equal. In the Poincar\'{e} sphere representation, a general
depolarizing channel is described by mapping the sphere surface
(where $D=1$) to a smaller contained ellipsoid surface. An isotropic
process is described by mapping the sphere surface to a smaller
concentric sphere.

Consider the Lyot depolarizer. If the angle between its two crystals
is tuned, depolarization is partial, but non-isotropic. The initial
polarization sphere is mapped onto an ellipsoid with two of the
radii equal zero (i.e., a line). By using two crystals of identical
length and tuning the angle between them, different channels are
produced \cite{Shaham,Shaham_QPT}. All of these channels are
non-isotropic, except at an angle of $54.7^\circ$, when the
polarization sphere is mapped onto another sphere of radius 1/3.

\begin{figure}[tbp]
\includegraphics[angle=0,width=60mm]{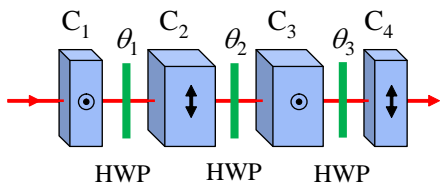}
\caption{\label{Iso_depo_configuration} (Color online) The isotropic
depolarizer: $C_1-C_4$ are the Calcite crystals, where the axes'
direction denote the fast axis orientation of each crystal.
$\theta_1$ and $\theta_3$ are the fixed half-wave plate (HWP)
angles. $\theta_2$ is the rotatable HWP angle, which controls the
amount of depolarization.}
\end{figure}

In this work, we study the depolarizing configuration of four fixed
birefringent crystals $(C_1,..,C_4)$ with three half-wave plates
(HWP) in between them (See Fig. \ref{Iso_depo_configuration}). The
lengths of $C_2$ and $C_3$ are twice the lengths of $C_1$ and $C_4$.
The fast axes of $C_1$ and $C_3$, and the slow axes of $C_2$ and
$C_4$ are parallel, and define the zero angle of the wave plates.
The first and the third HWP angles are related by
$\theta_3=-\theta_1$. The second HWP angle $\theta_2$ is tunable,
and controls the depolarization process.

\begin{figure}[tbp]
\includegraphics[angle=0,width=84mm]{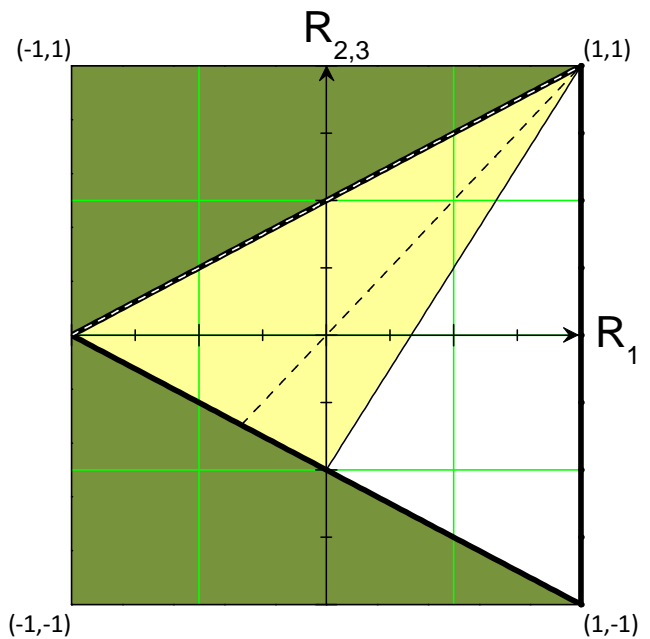}
\caption{\label{PossibleChannels} (Color online) Values of the radii
$R_1$ and $R_2=R_3$. The possible values of the presented
configuration are within the yellow zone. The isotropic channel is
represented by the black dashed line, the two crystal channel by the
white dashed line, and the dephasing channel by the $R_1=1$ line.}
\end{figure}

We have found that this configuration does not introduce additional
rotations, except for reflections along the $S_2$ and the $S_3$
axes. These reflections can be compensated for by an additional
half-wave plate at a zero angle before or after the setup. As there
are no additional rotations, the radii $R_i$ of the mapped
ellipsoids are aligned along the axes $\{S_i\}$ of the coordinate
system. The radii dependence on the HWP angles is:
\begin{eqnarray}\label{Radiuses}
\nonumber R_1&=&\cos^22\theta_2-\sin^22\theta_2\cos^24\theta_1\,, \\
\nonumber R_2&=&\cos^22\theta_2-\frac{1}{2}\sin^22\theta_2\sin^24\theta_1\,, \\
R_3&=&R_2\,.
\end{eqnarray}
Two radii are always equal. Negative values correspond to a
reflection along the corresponding radius direction. The possible
channels with this configuration are depicted in Fig.
\ref{PossibleChannels}. The requirement that the process matrix
should have eigenvalues between 0 and 1, that sum up to 1, dictates
that physically possible radii can not reside in the dark green
zones \cite{Karpinski,King01}. By examining the parameter space of
$\theta_1$ and $\theta_2$ it can be shown that our configuration can
create any channel within the yellow zone. When $\theta_2=0$, the
configuration has no effect on the initial state (all radii are
equal to 1). The configuration of two identical crystals, as
reported in Refs. \cite{Shaham} and \cite{Shaham_QPT}, creates
channels along the white dashed line. The configuration presented
here reproduces the two crystal scheme when $\theta_1=0$ or
$45^\circ$. The dephasing channel is represented by the line
$R_1=1$.

From Eqs. \ref{Radiuses} it can be shown that isotropic
depolarization (i.e., $D=R_1=R_2=R_3$) is achieved when the first
HWP is fixed at an angle of
$\theta_1=\tan^{-1}(\sqrt{2})/4\simeq13.68^{\circ}$, or at an angle
of $\theta_1=45^\circ-\tan^{-1}(\sqrt{2})/4\simeq31.32^{\circ}$.
These isotropic channels are represented by the black dashed line in
Fig. \ref{PossibleChannels}. The dependence of the degree of
polarization on the angle $\theta_2$ in this case is:
\begin{equation}\label{DOP}
D=\frac{1}{3}+\frac{2}{3}\cos 4\theta_2\,.
\end{equation}
For $\theta_2=0$, no depolarization occurs, while for
$\theta_2=30^\circ$, the depolarization is complete.

We built this depolarizer with $\theta_1\simeq31.32^{\circ}$ and
performed QPT for $\theta_2$ angles in the range
$0\leq\theta_2\leq45^\circ$. Photon pairs were generated by
spontaneous parametric down-conversion of 390\,nm pulses. One photon
of the pair was probabilistically split by a beam splitter, and sent
to a single-photon detector (SPD). The second photon was sent to the
depolarizer. Before entering the depolarizing unit, the photons were
spectrally filtered by a 5\,nm band-pass filter and spatially
filtered by coupling them into a single-mode fiber, and then
collimated into free space. The photons were prepared in the
polarization states $|H\rangle$, $|V\rangle$, $|P\rangle$, and
$|R\rangle$, which served as the initial states for the QPT
procedure. The depolarizer was made of Calcite crystals that were
2\,mm and 1\,mm long. After depolarization, the final polarization
state was characterized at the QST unit which was composed of half-
and quarter-wave plates, a polarizer and another SPD. Detecting the
two photons in coincidence reduced the background counts.

\begin{figure}[tbp]
\includegraphics[angle=0,width=84mm]{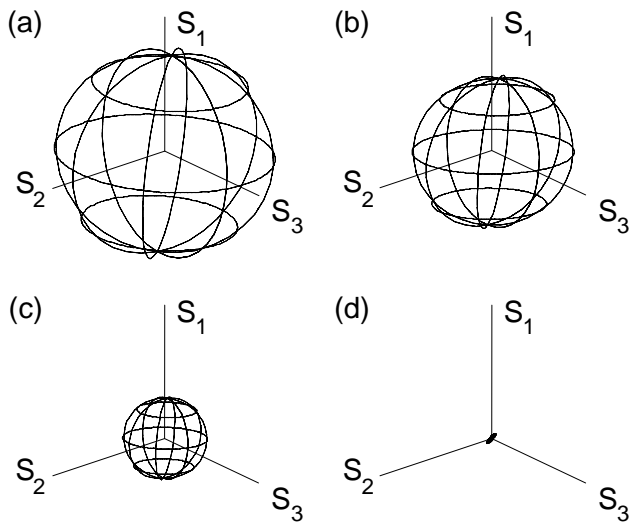}
\caption{\label{isotropic_processes}(Color online) Experimentally
measured processes in the Poincar\'{e} sphere representation. The
depolarization maps correspond to $\theta_2$ rotations of the second
HWP at (a) $4^\circ$, (b) $15^\circ$, (c) $22^\circ$, and (d)
$30^\circ$.}
\end{figure}

Experimentally measured processes at four $\theta_2$ angles are
shown in Fig. \ref{isotropic_processes}. The four measured maps in
the Poincar\'{e} sphere representation, were taken at $\theta_2$
values of $4^\circ, 15^\circ, 22^\circ$ and $30^\circ$, which
correspond to theoretical final values of $D$ of 0.97, 2/3, 0.36,
and 0, respectively. All processes were reconstructed using the same
maximal likelihood protocol that was previously used in Ref.
\cite{Shaham_QPT}, in order to restrict their parameters to
physically allowed values.

\begin{figure}[tbp]
\includegraphics[angle=0,width=84mm]{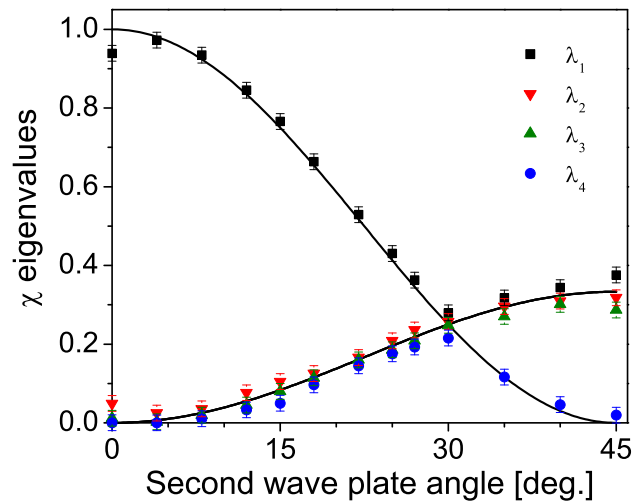}
\caption{\label{Chi_eig_results}(Color online) The eigenvalues
$\lambda_1-\lambda_4$ of the $\chi$ matrix as a function of the
$\theta_2$ angle of the second half-wave plate. Theoretical
predictions are presented as solid lines.}
\end{figure}

In order to test the isotropy of the constructed channels, the
eigenvalues $\lambda_i$ of the process matrix were studied. Figure
\ref{Chi_eig_results} shows the measured eigenvalues of the
reconstructed four-dimensional $\chi$ matrices as a function of the
second HWP angle $\theta_2$. All the measured eigenvalues are in
good agreement with their theoretical predictions. In particular,
three of the four eigenvalues have similar values. For $\theta_2$
angles close to zero, which represent processes of almost no
depolarization, the highest eigenvalue deviates the most from
theory. This is a result of the maximal likelihood procedure
correction for negative eigenvalues which should be zero or
positive-but-small.

Controllable isotropic depolarization can also be realized with
other setup parameters. The main constraint on the respective
crystal lengths $L_1,..,L_4$ is that $L_1=L_4$ and $L_2=L_3$ in
order for zero depolarization to be possible. Surprisingly, the
ratio $L_2/L_1$ can be any number (even less than 1) except for
exactly 1 or 1/2. The demand for full temporal mode separation
requires that $L_{min}=\textrm{min}(L_1, L_2, |L_1-L_2|)$ satisfies
Eq. \ref{temp_delay}. If the light has a relatively long coherence
length, the birefringent crystals can be replaced by polarization
preserving birefringent fibers, which can induce much higher
temporal separation due to their length.

The depolarizer can even be built without any wave plates. The
effect of a wave plate that is oriented in a certain angle $\theta$
can also be achieved by rotating all proceeding elements by an angle
of $2\theta$. Thus, setting the two middle crystals $C_2$ and $C_3$
at an angle of $27.4^\circ$ or $62.6^\circ$ relative to the
configuration of Fig. \ref{Iso_depo_configuration} is equivalent to
the zero depolarization setting. Then, the amount of depolarization
is controlled by applying a relative rotation between the left
($C_1$ and $C_2$) and the right ($C_3$ and $C_4$) pairs of crystals.
Complete depolarization is reached when this rotation is set to
$60^\circ$. Note that such a configuration will result in additional
polarization rotations.

In conclusion, we have described and demonstrated a controlled
isotropic depolarizer. The depolarizer reduces the degree of
polarization of any input polarization state by the same extent, to
any required level. The depolarizer is composed of four fixed
birefringent crystals and three half-wave plates. It is suitable for
light of short coherence length. The induced depolarization was
characterized using a quantum process tomography procedure. High
fidelity with the theoretical predictions was observed. This
depolarizer can also be realized without wave plates and the
crystals can be replaced by birefringent fibers for light with
longer coherence length. A possible application of this depolarizer
is the realization of controlled quantum noise in photonic quantum
communication channels in order to test quantum communication
protocols.

We would like to thank the Israeli Ministry of Science and
Technology for financial support.

\end{document}